# Optimizing Waste Management Collection Routes in Urban Haiti: A Collaboration between DataKind and SOIL

Michael Dowd
DataKind
Brooklyn, NY, USA
michael@datakind.org

Anna Dixon
DataKind
Washington, D.C., USA
anna@datafoss.com

Benjamin Kinsella
DataKind
Brooklyn, NY, USA
benjamin.kinsella@datakind.org

**ABSTRACT**
Sustainable Organic Integrated Livelihoods (SOIL) is a research and development organization that aims to increase access to cost-effective household sanitation services in urban communities in Haiti. Each week, SOIL provides over 1,000 households with ecological sanitation toilets, then transports the waste to be transformed into rich compost. However, SOIL faces several challenges regarding the route optimization of their mixed fleet vehicle routing. This paper builds upon the authors' submission to Bloomberg's 2019 Data for Good Exchange (D4GX), presenting preliminary findings from a joint collaboration between DataKind, a data science nonprofit, and SOIL. This research showcases how optimization algorithms and open source tools (i.e., OpenStreetMap and Google OR-Tools) can help improve - and reduce the costs - of mixed fleet routing problems, particularly in the context of developing countries. As a result of this work, SOIL is able to make improvements to their collection routes, which account for different road conditions and vehicle types. These improvements reduce operational costs and fuel use, which are essential to the service's expansion in the coming years.

## 1. INTRODUCTION
Waste management and sanitation pose significant challenges in many developing countries, where approximately 36% of the global population lacks access to a toilet (UNICEF/WHO, 2015). More alarmingly, human waste, which is often dumped directly into waterways or left untreated, has been shown to permeate groundwater and contribute to health and environmental problems (e.g., Baum and Bartram, 2013). This issue is particularly evident in Haiti, as exposed in Walker et al.'s (2012) study, where childhood diarrheal incidence rate and cholera epidemics are among the highest in the world. That is, only 30% of the population has access to improved sanitation and 1% of wastes are safely treated, contributing to one of the largest cholera epidemics in the world. According to a report conducted by the Pan American Health Organization and World Health Organization, a total of 13,681 cases of cholera and 159 deaths were reported in Haiti in 2017 (PAHO/WHO, 2018).

In response to this problem, SOIL, a nonprofit working in Haiti, increases access to cost-effective household sanitation services in urban communities. The nonprofit provides over 1,000 households with container-based sanitation (CBS) toilets. SOIL safely produces more than 40 metric tons of agricultural grade compost each month that is used throughout the country.

Despite SOIL's expansion, however, the organization faces several challenges regarding the optimization of their vehicle routes. The first concerns the logistics and cost of managing their fleet of vehicles, each with limited capacity. Put differently, each vehicle must visit a group of customers in the most efficient and cost-effective manner. Prior to this project, SOIL determined routes and customer distribution based on local knowledge, experience and iteration. A second challenge the organization faces concerns the lack of affordable optimization tools and algorithms, adding additional costs to the organization. Turning to this project, there was an opportunity to leverage open source tools to develop new custom software and tuning that could be iteratively customized for SOIL's context. Since vehicle transportation and its associated costs are the organization's biggest cost-driver, improving the efficiency of this process could help SOIL provide dignified sanitation to more people. SOIL hoped to find a better and faster solution for generating this information.

To address these problems, SOIL partnered with DataKind, a global nonprofit that harnesses the power of data science and AI in the service of humanity, in order to examine how optimization algorithms can help improve - and reduce costs - of sanitation services. With a team of expert data scientists, the project was guided by two overarching questions: (1) What is the fewest number of trips needed to service all of SOIL's customers when the number of containers might vary from house to house? (2) What is the most efficient route for each of those trips?

This paper is organized as follows. Below, a general introduction to optimization algorithms is provided, namely several Vehicle Routing Problems (VRPs), followed by the project's methods that includes data preparation, optimization of model, and its visualization. Next, the output is revealed, which is then summarized and discussed.

## 2. THE CHALLENGE

To approach the challenge of route optimization, this project adapted a variant of the VRP originally outlined by Dantzig and Ramser (1959). VRP is among the most studied combinatorial optimization problems, which according to Baldacci, Battarra and Vigo (2008), consists of the optimal design of routes to be used by a fleet of vehicles to serve a set of customers. Since its original inception, current models give greater attention to more complex variants of VRP, including the Capacitated VRP (CVRP) and VRP with Time Windows (VRPTW) (see Baldacci et al., 2008 for overview).

This project focuses on "rich" VRPs, namely Mixed Fleet or heterogeneous fleet VRPs, which are closer to practical distribution problems observed in the field (e.g., Bräysy et al. 2002; Cordeau et al., 2006; Toth & Vigo, 2002). This variant is characterized by vehicles having different capacities and costs, which are available for distribution to multiple depots and trips. In the context of this project, SOIL operates a fleet of vehicles each with limited capacity, which make visits to segments of customers every week. These vehicles are then returned to the depot where they are unloaded and made available for another route.

To exemplify this process, consider Figure 1, which demonstrates vehicle routes in loading and unloading waste between the depot and customer.

**Figure 1: Visualization of rich VRPs based on SOIL vehicle route**

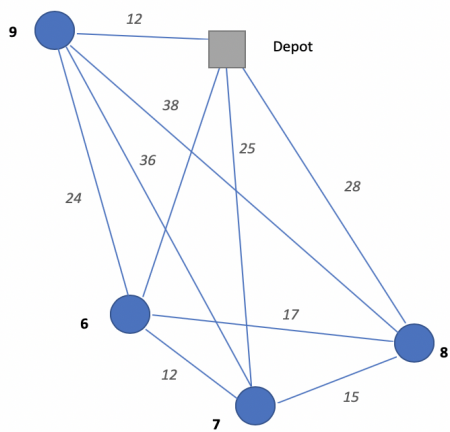

The above figure graphically represents the challenge faced by SOIL. Each blue marker is called a node, representing an individual customer. The values adjacent to each node represent the number of buckets collected by SOIL vehicles. Next, the edges, namely the lines connecting each node, depict the routes between customers. The values along each edge represents the distance, or cost, between each node. Thus, as a rich VRP problem, SOIL was faced with an optimization challenge and the following questions. Given a fleet of vehicles with limited capacity, what is the fewest number of trips needed to service all of SOIL's customers when the number of containers might vary from house to house. Furthermore, how can these efficient routes be affordably mapped using open-source software?

In the next section, details on the project's methods are provided, followed by the open-source software used to create the final product.

## 3. METHODS

To begin, the team cleaned and prepared datasets for optimization. These data included (1) information on individual customers; (2) SOIL facility locations; and (3) vehicles. Descriptions of these data are shown in Table 1:

| Table | Description |
| --- | --- |
| Customer | GPS records for each household, length of time with service, number of toilets on contract, number of containers expected - when applicable - and their service zone. |
| SOIL Facility Locations | GPS records for all locations required by the tool and not included in the customer file (e.g., the depot or focal points) |
| Vehicles | Records of vehicle types and the different rules on how fast they can go on different roads; What roads vehicles can or cannot access. |

### 3.1. MODELING AND OPTIMIZATION

After cleaning, the team then prepared the data to be used for modeling and optimization. This process included the calculation of distances and times between nodes and edges, notably the SOIL vehicle routes needed to account for local road conditions. To do so, the team leveraged two open-source software. The first, OpenStreetMap (OSM), provided the most up-to-date information on the local roads for the SOIL locations. And the second, Open Source Routing Machine (OSRM), a routing software that computes distances and times between points based on the OSM data, formulated the routes.

It is important to note that OSRM is a library that performs operations on any network described in the OSM format.

Therefore, custom software was written to simplify the running of the OSRM engine, including Python bindings to the OSRM engine, leveraging the native C++ library directly rather than through HTTP queries to separate server processes hosting an instance of the OSRM engine tuned to a single vehicle profile. This involved minor changes to the OSRM API source and the creation of a custom set of functions in C++ exposed to our Python application with a single importable module, where the interoperability between Python and C++ was provided by pybind11. The bindings enabled us to manage multiple vehicle profiles efficiently (both in memory usage and development effort), avoided the complexity of executing multiple web servers in the background, removed the need of internal network calls, and resulted in approximately a 10X computation time improvement. The solution used the OSM maps for Haiti and leveraged OSRM to produce matrices of distances and travel times between the customers and the detailed segments of the routes.

Relevant to the rich VRP problem on hand, SOIL had numerous vehicle types with different capacities and dimensions. Most of SOIL's operations are carried out by three-wheelers and wheelbarrows. Three-wheelers are the most commonly used vehicles as they handle certain terrains better than larger vehicles, such as smaller roads that are inaccessible, but still allow fast and efficient travel. Wheelbarrows that are slower and larger are required in certain regions, particularly in areas where vehicles are not permitted, and customers are densely populated. To account for these differences, the team produced time and distance matrices for each vehicle type, which served as an input to the optimization. The team accomplished these matrices by creating vehicle-specific profiles. Vehicle profiles define various parameters such as average speeds and what roads and paths are available to specific vehicles. Allowable roads are indicated in the profile using metadata about the actual OSM link data as indicated by OSM tags.

Next, we turn to the optimization of route assignments. The goal was to find round-trip routes for each vehicle, which constituted the departure from - and termination at - the depot, making multiple visits to a list of customers in a prescribed order. To account for the difficulty of vehicle route optimization, the team used Google OR-Tools, free open-source software that facilitates in the vehicle routing, flows, integer and linear programming, and constraint programming. To utilize Google OR-tools a user must possess conceptual and technical expertise as well as the ability to build custom code/software to interface with the library. The team developed the necessary software to feed the required inputs into Google OR-tools to minimize the total travel time to cover all the households. The core inputs were (1) the coordinates of each household and depot; (2) the buckets per household; (3) the capacity of each vehicle type; and (4) the travel time and distance matrices between each depot and every household, as well as from every household to every other household, for each type of vehicle. The team generated software fed this into a capacitated vehicle routing problem in the format expected by Google-OR Tools to produce the order of customer visits.

The optimization component presented several challenges. Before the model and optimization solver tuning phase, the team's solutions would often take visually obvious routes, which were not optimized. We corrected many of these issues by addressing the cost function scaling and local search algorithm. Initially, when we set the cost function for minimizing travel time, customers that were close together did not always seem to be in the correct order. The team found that the time matrices were originally constructed in minutes. However, Google OR-Tools routing solver only works on integer cost functions. As a result, the model's perspective of customers who were less than one minute from each other - a common occurrence for SOIL - was that they were indistinguishable. The cost function units were then changed to seconds to avoid rounding errors and improve performance. Furthermore, the performance was improved by experimenting with the different local search algorithms available to Google OR-Tools. That is, local search algorithms allow the solver to escape local optima, such that the guided local search algorithm was more appropriate than Google OR-Tools automatic selection.

The final optimization challenge arose from "out-and-back" roads. In these roads, vehicles would enter to visit a group of customers. They then would turn around after the last customer and exit through the same entrance in which they entered, as opposed to continuing to the end and exiting on the other side. A challenge was that the solver's solution did not follow the common sense of collecting neighboring customers together. The solution would suggest a vehicle stop at a cluster of customers to collect from a subset of that group and travel to the end of the road, while revisiting that cluster's remaining customers on the way back. An example of this scenario is shown in Figure 2:

**Figure 2: Solution stops for customers**

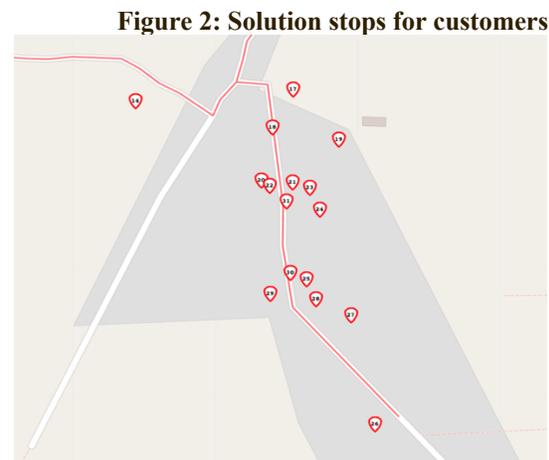

As shown in Figure 2, the solution stops at customers 20 through 24, proceeding down the road to collect customers 25 through 29. It would then pick up customers 30 and 31 on the way back, despite these customers being very close to the other clusters. This challenge would occur because the solver does not have a concept of the cost of the driver getting on and off the vehicle, and any order of the customers on an out-and-back road is mathematically equivalent. Addressing this issue, a clustering algorithm was implemented whereby the solver would be slightly penalized for not collecting very close customers once stopped.

The last step the team took was to transform the optimized routes into a format that the SOIL team could actually use on the ground. The team created interactive web maps using Python, OSRM and the open-source library, Folium. The team output "master" maps with all routes and all customer zones, as well as maps by zone that SOIL collectors could load on their mobile devices. Each map showed each round trip that each vehicle needed to complete to visit all customers. The expected workflow is that a SOIL dispatcher will load the relevant map on the mobile device of the collector and the collector will use the maps in the field to see their routes and the sequence of customers to visit.

To summarize, three steps were taken to transform data into optimized routes, which are reviewed in Figure 3:

**Figure 3: Data processing steps**

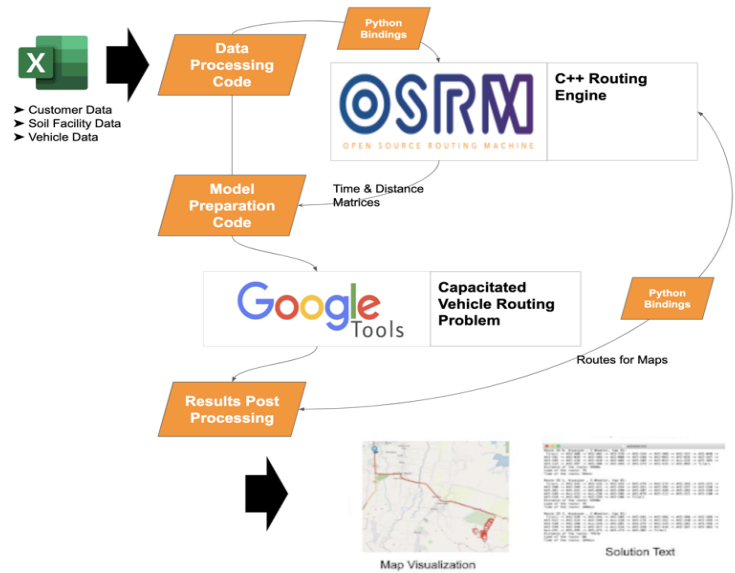

The above figure outlines the steps taken to transform data into optimized routes. These steps included data preparation, modeling and optimization of vehicle routes, and visualization. Having described the preparation and optimization of these data, we now turn to present the output of the above optimization models.

## 4. MODEL OUTPUT

This section presents case examples of the optimization output, which are reviewed as optimized routes in select service zones. The software outputs a variety of maps for both SOIL administrators and collectors. These maps provide three pieces of information: (1) the routes, (2) the customers to visit, and (3) some metadata on customers in case a collector needs to contact a customer. Naive comparisons showed that prior to using the model the SOIL collectors had a total of 38 trips to visit all customers, while the model provided a total of about 33 trips to visit all customers leading to about a 13% decrease in trips. Below are some example outputs, shown in Figures 4 and 5:

**Figure 4: Master map containing all optimized routes in multiple service zones**

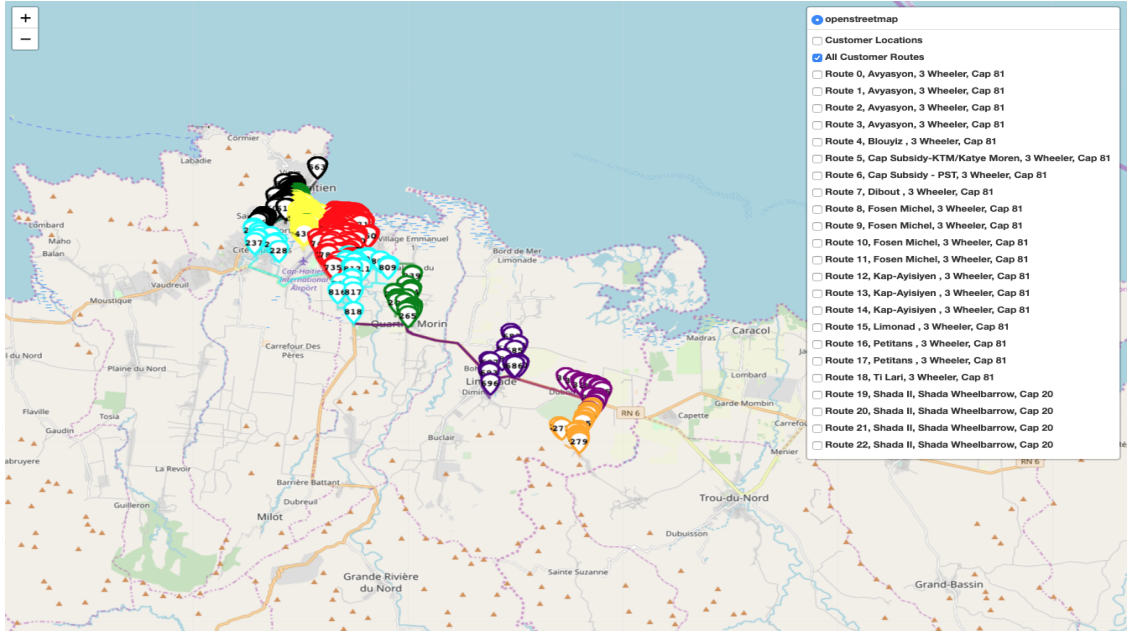

*Note: Different colors indicate different service zones.*

**Figure 5: Example of optimized routes for one service zone**

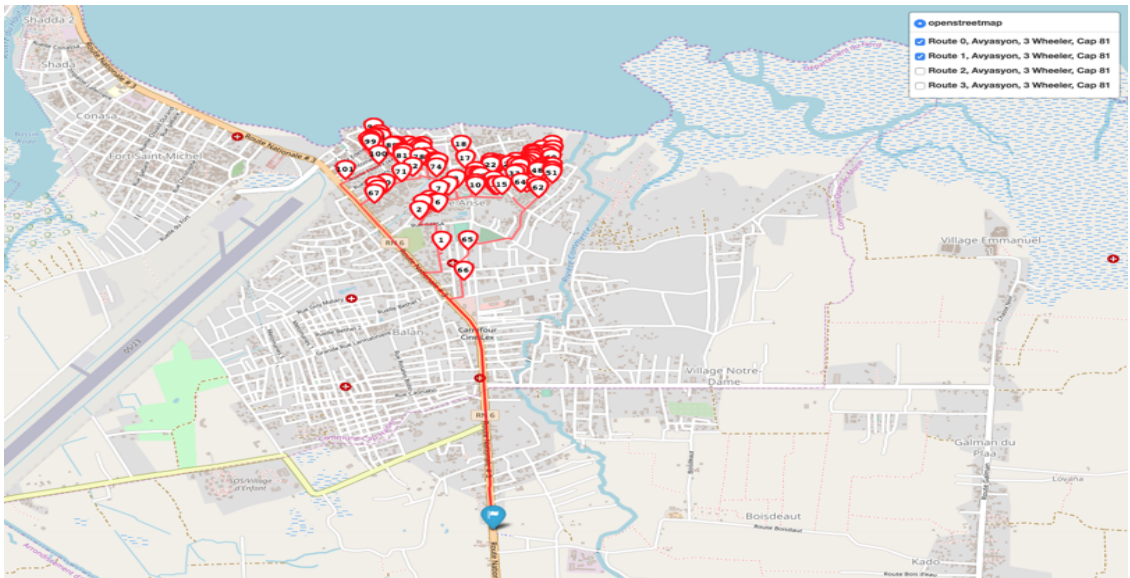

*Note: Each route, selectable in the key, is a driver's round trip from the depot to a group of customers.*

First, in Figure 4, we observe that SOIL utilizes service zones that roughly correspond to different neighborhoods, such that the model was run by service zone. Vehicle routes for every customer in many of the service zones in which SOIL operates. The map depicts the routes and customer sequencing for visiting all customers while maintaining vehicle capacities for seven zones. Colors represent different zones. By using the layer button in the top right corner of the map, different vehicle trips can be activated/deactivated.

Next, Figure 5 shows an example of an optimized route in one service zone, namely Avyasyon, a neighborhood of Cap-Haïtien in Port-au-Prince. The zone route map files include all routes within the Avyasyon zone. As with the master map, different routes can be visualized by using the layer button in the top right corner. For each route, the starting and ending point is the depot, which is indicated with a blue marker with a flag. All other markers indicate customer stops and are numbered in the tool's suggested visited order (e.g., "1" should be the first customer, etc.). Clicking on the customer marker creates a pop-up with the customer ID and phone number.

## 5. CONCLUSION

This paper presented results from a collaboration between DataKind and SOIL, showcasing how open-source software can be used to generate optimized vehicle routes. To summarize, SOIL, which operates in Haiti, addresses concerns over waste management and sanitation, by providing households with container-based sanitation (CBS) toilets and transforming all collected waste in agricultural-grade compost. While the nonprofit serves over 1,000 homes, SOIL faces several logistical constraints that contribute to transportation costs and limits their expansion. More specifically, there is a fleet of vehicles each with limited capacity and must visit a group of customers each week in the most efficient and cost-effective manner. SOIL only used ad hoc methods for creating routes and desired to have a more consistent and standardized approach, ideally harnessing optimization techniques.

To address this problem, a team of pro bono data scientists leveraged custom software, OpenStreetMap and Google OR-Tools to generate optimized vehicle routes that accounted for different road conditions and different vehicle types. These optimized routes were then exported to simple web maps that drivers can reference while traveling. As a result of these tools, SOIL has been able to make improvements to their collection routes, which are anticipated to reduce costs and fuel use. Such efficiencies are essential to the service's expansion in the coming years.

The outcome of this project has potential for long-lasting impact. First, improvements to SOIL's collection routes are likely to reduce costs and fuel use into the future. Second, this project exemplifies the use of open-source software as a way to not only lower-costs associated with transportation and logistical constraints, but also expensive software licenses. The tools that were employed in this project are free and, with additional custom software, provide a reasonable solution to SOIL's need.

## 6. ACKNOWLEDGEMENTS

We would like to thank the team of talented data scientists on this project, including Annabel Buckfire, Anna Dixon, Shubin Li, Nikolaj van Omme, Conor Owen, Sebastian Ouellet, and Rachael Rho. We would also like to thank our project partners at SOIL for their guidance and subject matter expertise, specifically Erica Lloyd, Jess Laporte, and Djifferson Sainfort. Furthermore, this paper and project was supported by a grant from the 11th Hour Project.